\documentclass[12pt]{article}
\usepackage{amsmath,amssymb,epsfig}
\begin{document}
\parindent=0pt
\parskip=6pt
\rm

\vspace*{0.5cm}

\begin{center}
{\bf THERMODYNAMIC PROPERTIES OF THE PHASE TRANSITION TO
SUPERCONDUCTING STATE IN THIN FILMS OF TYPE I SUPERCONDUCTORS}

\vspace{0.3cm}
 D. V. SHOPOVA$^{\ast}$, T. P. TODOROV$^{\dag}$,  and D. I. UZUNOV$^{\ddag}$

{\em  CPCM Laboratory, G. Nadjakov Institute of Solid State Physics,\\
 Bulgarian Academy of Sciences, BG-1784 Sofia, Bulgaria.} \\
\end{center}

$^{\ast}$ Corresponding author: sho@issp.bas.bg

$^{\dag}$ Permanent address: Joint Technical College at the Technical
University of Sofia.

$^{\ddag}$ Temporal address: Max-Planck-Institut f\"{u}r Physik komplexer Systeme,\\
N\"{o}thnitzer Str. 38, 01187 Dresden, Germany.

\vspace{1cm}


\vspace{0.2cm}


\vspace{0.2cm}
\begin{abstract}
The effect of magnetic fluctuations on the free energy, the order
parameter profile and the latent heat at the equilibrium point of the
first order phase transition to superconducting state in thin films of
type I superconductors is considered. The possibility for an
experimental observation of the fluctuation change of the order of the
superconducting phase transition is briefly discussed. Numerical data
for the order parameter jump and the latent heat of Al films are
presented for needs of experimental studies.
\end{abstract}

{\bf 1. Introduction}

The investigation of the fluctuation-induced weakly-first order phase
transition in type I superconductors in a zero magnetic field known as
Halperin-Lubensky-Ma (HLM) effect~\cite{HLM:1974} has been recently
extended to the case of thin superconducting films~\cite{FSU:2001}. It
has been shown~\cite{FSU:2001} that the HLM effect in
quasi-two-dimensional (quasi-2D) films is much stronger than in bulk
(3D) systems~\cite{HLM:1974, TET:2002}. This result opens an
opportunity for an experimental observation of the effect in suitably
chosen superconducting films.

In this letter we present new and more precise results for the
behaviour of the free energy, the order parameter, and the latent heat
at the equilibrium phase transition temperature for thin films of type
I superconductors. The numerical values of physical quantities of
experimental interest are calculated for Al films. Our results are
compared to those given in Refs.~\cite{HLM:1974,FSU:2001, TET:2002,
CLN:1978}. A detailed information about the HLM effect of a fluctuation
change of the order of the superconducting phase transition in a zero
mean magnetic field due to persisting magnetic fluctuations and the
methods of investigation of this phenomenon are published in
Refs.~\cite{CLN:1978,FH:1999,UZ:1993}. We shall follow the notations
from Ref.~\cite{LP:1980} for the parameters of the Ginzburg-Landau (GL)
free energy of superconductors.

{\bf 2. Effective free energy}

The starting point of our consideration is the effective free energy
density $f(\psi) = F(\psi)/V$ of a type I superconductor with a volume
$V$ which is a function of the mean (uniform) superconducting order
parameter $\psi = <\psi(\vec{x})>$ of the form (see, e.g.,
Ref.~\cite{FSU:2001}):
\begin{equation}
f(\psi) = f_0(\psi) + \delta f(\psi)\:,
\end{equation}
where
\begin{equation}
f_0(\psi) = a|\psi|^2 + \frac{b}{2}|\psi|^4\:,
\end{equation}
and
\begin{equation}
\delta f(\psi) = \frac{1}{2}(D-1)k_BT\sum_{\vec{k}}^\Lambda
\mbox{ln}\left[1+ \frac{\rho(\psi_0)}{k^2}\right]\:.
\end{equation}
In Eqs.~(2)~-~(3), $\rho(\psi_0) = \rho_0|\psi|^2$, where $\rho_0 =
(8\pi e^2/mc^2)$, $a = \alpha_0(T-T_{c0})$ and $b > 0$ are the usual
Landau parameters. They are related to the zero temperature coherence
length $\xi_0 = (\hbar^2/4m\alpha_0T_{c0})^{1/2}$, the zero-temperature
critical magnetic field $H_{c0} = \alpha_0T_{c0}(4\pi/b)^{1/2}$, and
the initial (unrenormalized) critical temperature
$T_{c0}$~\cite{LP:1980}. The term $\delta f(\psi)$ in $f(\psi)$
describes the effect of the magnetic fluctuations. In Eq.~(3), the sum
over the wave vector $\vec{k}$ is truncated by the upper cutoff
$\Lambda \geq k \equiv |\vec{k}|$. For films of thickness $L_0$ and
volume $V = (L_0L_1L_2)$ we shall assume periodic boundary conditions,
and $a_0 \ll L_0 \leq \Lambda^{-1} \ll L_{\alpha}$; $\alpha = 1,2$. For
quasi-2D film, i.e. films obeying the condition $a_0 \ll L_0 \leq
\Lambda^{-1}$, the sum in Eq.~(1) contains only terms with zero
component $k_0 = 2\pi n_0/L_0 (n_0 = 0)$ of the wave vector $\vec{k} =
(k_0, k_1,k_2)$. This means that the thickness $L_0$ should be smaller
than the magnetic penetration depth $\lambda (T) =
(\lambda_0/|t_0|^{1/2})$ which gives the characteristic length of the
magnetic fluctuations; $\lambda(0) \equiv \lambda_0 =
(b/\rho_0\alpha_0T_{c0})^{1/2}$ is the zero-temperature penetration
depth, and $t_0=(T-T_{c0})/T_{c0}$. We choose the cutoff
 $\Lambda = (\pi/\xi_0)$, so we shall study  films of size $L_0 <
\xi_0$. This is consistent with the general requirement $\xi_0 <
\lambda(T)$ for the validity of the GL free energy for the type I
superconductors (see, e.g., Ref.~\cite{LP:1980}).

For quasi-2D systems, the continuum limit applied to the sum in Eq.~(3)
yields a simple integral over $\vec{k} = (0,k_1,k_2)$. Solving this
integral we obtain
\begin{equation}
\delta f(\psi) = \frac{k_BT}{4\pi L_0}\Lambda^2 \left[ \left(1 +
\frac{\rho_0|\psi|^2}{\Lambda^2}\right) \mbox{ln}\left(1 +
\frac{\rho_0|\psi|^2}{\Lambda^2}\right) -
 \frac{\rho_0|\psi|^2}{\Lambda^2}\mbox{ln}\left(\frac{\rho_0|\psi|^2}{\Lambda^2}
  \right)\right]\:.
\end{equation}
The second term in the r.h.s. of Eq.~(4) is nonanalytical and cannot be
expanded in powers of $\psi$. An incomplete Landau expansion of this
free energy in positive  powers of $|\psi|^2$ can be performed,
provided
\begin{equation}
(\rho_0|\psi|^2/\Lambda^2) \ll 1\;.
\end{equation}
This inequality should be satisfied in the stable (Meissner) phase
$\psi(T) >0$, i.e., at the absolute minimum $f[\psi(T)]$ of the
function $f(\psi)$ below the phase transition temperature.

Hereafter we shall denote by $\psi(T)$ the equilibrium value of $\psi$,
which describes a stable (or metastable) phase and is a solution of the
equation of state $(\partial f/\partial \psi) = 0$. The condition~(5)
and the problems included in the reminder of this article can be
considered in terms of auxiliary parameters corresponding to the
theory, in which the $\delta f(\psi)$-part of the free energy is
neglected, i.e., corresponding to the free energy $f_0(\psi)$ given by
Eq.~(2). Within this simplified theory the modulus of the equilibrium
zero-temperature order parameter is given by $|\psi_0| = |\psi(T=0)| =
(\alpha_0T_{c0}/b)^{1/2}$. This quantity can be used to define the
reduced order parameter $\varphi = (|\psi|/|\psi_0|) \geq 0$ of the
general theory represented by the complete free energy $f(\psi)$. At
equilibrium, i.e., $\psi = \psi(T)$,  the condition (5) can be written
in the form $[\varphi(T)/\lambda_0\Lambda] < 1$. If we suppose that the
difference between the values of $\varphi(T)$ for the simplified and
complete theories can be ignored, we shall obtain that $\lambda
(T)\Lambda > 1$, which is consistent within the superconductivity
theory; see our discussion below Eq.~(3). We have recently investigated
Eq.~(4) with the help of the Landau expansion in Ref.~\cite{FSU:2001}.
Here we shall consider the free energy $f(\psi)$ given by
Eqs.~(1),~(2), and~(4) without such an expansion.

Using the reduced order parameter $\varphi$ the free energy $f(\psi)$
can be represented in the form
\begin{equation}
f(\varphi) =\frac{H^2_{c0}}{8\pi}\left\{2t_0\varphi^2 + \varphi^4 + C(1
+
t_0)\left[\left(1+\mu\varphi^2\right)\mbox{ln}\left(1+\mu\varphi^2\right)
- \mu\varphi^2\mbox{ln}\left(\mu\varphi^2\right)\right]\right\}\:,
\end{equation}
where
\begin{equation}
C= \frac{2\pi^2k_BT_{c0}}{L_0\xi_o^2H^2_{c0}}\:,
\end{equation}
and $\mu = (\xi_0/\pi\lambda_0)^2$.

{\bf 3. Results and discussion}

It is known~\cite{HLM:1974} that the HLM effect is stronger in type I
superconductors with relatively small GL parameter $\kappa =
(\lambda_0/\xi_0)$. This is obvious from Eq.~(6). That is why we can
choose as adequate examples the aluminium (AL, $\kappa \sim 0.01$,
$\xi_0 = 1.6\mu$m, $T_c \approx 1.19$K, $H_c(0) \approx 99 $Oe) and
tungsten (W, $\kappa \sim 0.001$, $\xi_0 = 37\mu$m, $T_c \approx 5$ mK,
$H_c(0) \approx 1.15$ Oe). One cannot be certain about the best choice
of the substance for an experimental check of the HLM effect before a
comprehensive consideration of the experimental problem.

Our concrete aim is to establish the size of the effect and we shall do
the calculations  for Al, which is probably one of the best candidates
for  experiments. The numerical values of the parameters $\tilde{C} =
CL_0$ and $\mu$, where $\lambda_0 = (\hbar c/2\sqrt{2}eH_{c0}\xi_0)$,
can be calculated from the experimental data for  $T_c$, $H_c(0)$, and
$\xi_0$ of Al, given above. Note, that the available experimental data
vary depending on the way of preparation of samples and the type of
measurement. Besides, the experimental values used here are for bulk
monocrystals of Al and differ within 10 - 20\% from those for thin Al
films of thickness much smaller than the value $L_0 = 0.1\mu$m
considered below. However, the variation of the experimental data, used
here does not essentially influence our numerical results.

\begin{figure}
\begin{center}
\epsfig{file=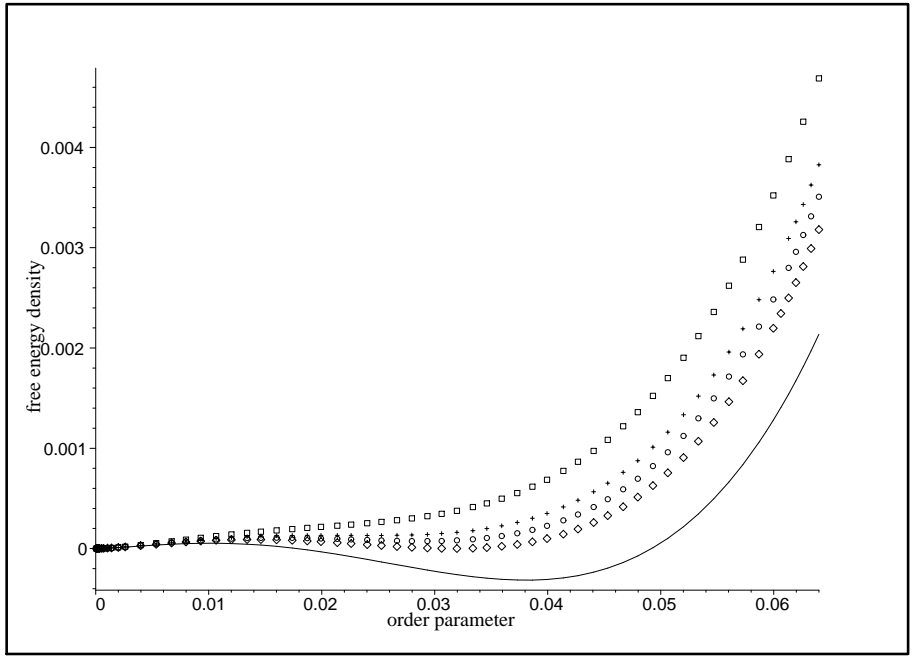,angle=-90, width=12cm}\\
\end{center}
\caption{Curves representing the free energy (6) for five values of
$t_0$: $t_0= - 0.001$ (see $\square$-line), $t_0= - 0.00127$ ($+$),
$t_0 = -0.00137$ ($\circ$), $t_0 = - 0.001473$ ($\diamond$), $t_0 = -
0.0018$ (---).} \label{dimo2f1.fig}
\end{figure}

\begin{figure}
\begin{center}
\epsfig{file=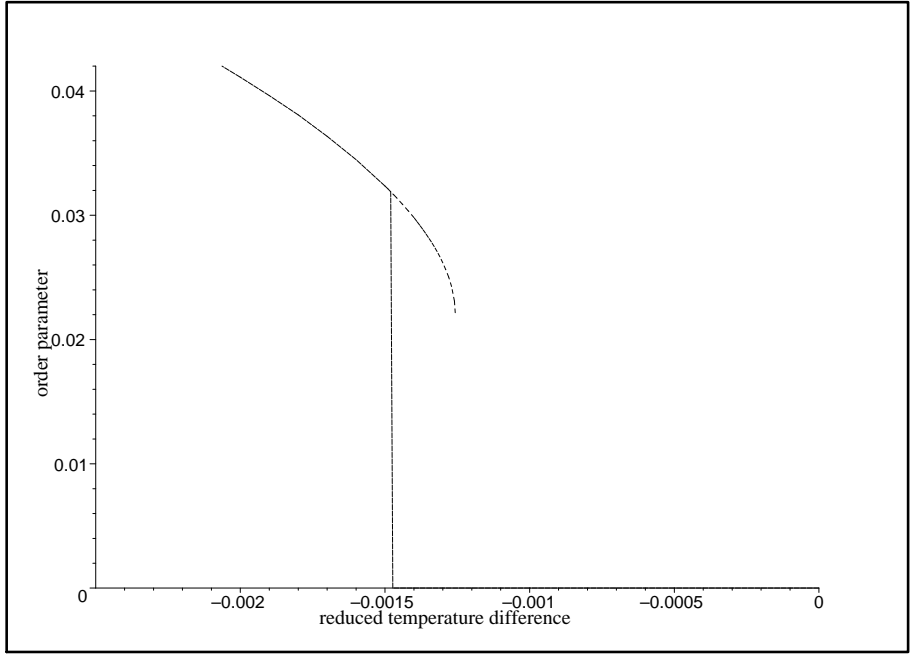,angle=-90, width=12cm}\\
\end{center}
\caption{Order parameter profile near $T_{c0}$. The vertical line at $t
= - 0.001473$ indicates the equilibrium jump of the order parameter.}
\label{dimo2f2.fig}
\end{figure}

The free energy density $f(\varphi)$ for Al films of thickness $L_0 =
0.1\mu$m and several values of the parameter $t_0$ is calculated from
Eq.~(4). The result is shown in Fig.~1. The reduced order parameter
corresponding to metastable and stable superconducting states is
calculated with the help of the equations $f(\varphi) = 0$,
 $(\partial f/\partial \varphi) = 0$, and the stability condition
 $\partial^2 f/\partial \varphi^2 > 0$ (see Fig.~2).

Fig.~1 exhibits a well established phase transition of first order.
This confirms the preceding results~\cite{FSU:2001} obtained with the
help of Landau expansion; note that in Ref.~\cite{FSU:2001} another
normalization of the effective free energy has been used, which seems
to be less convenient than the present one. The positive minima of the
free energy describe metastable superconducting states (overheated
superconductivity). The respective metastable values of the reduced
order parameter $\varphi$ are shown by the dashed curve in Fig.~2. The
metastability states are closed in the temperature interval from
$T^{\ast} = 0.9990T_{c0}$ to $T_{eq} \approx 0.9985 T_{c0}$ -- the
equilibrium temperature of the first order phase transition. The curve
marked by squares ($\square$) in Fig.~1 corresponds to $T=T^{\ast}$,
i.e., this is the curve, at which the minimum of the free energy for a
nonzero value of $\varphi$ appears for the first time when the
temperature is lowered from the side of the normal phase. The curve
drawn by "diamonds"($\diamond$) corresponds to the equilibrium
transition temperature $T_{eq}$, at which the minimum of the free
energy for nonzero value of the order parameter $\varphi$ is equal to
zero. There the energies of the superconducting and the normal phases
are equal. The superconducting phase is stable for all temperatures
below $T_{eq}$. The stable superconducting states are shown by the
solid curve in Fig.~2 for several values of $t_0$ below $t_{0eq} =
t_0(T_{eq}) \approx - 0.0015$.

The equilibrium entropy jump $\delta s = - df[T,\varphi(T)]/dT$ at
$T_{eq}$ per unit volume (the total entropy jump is $\delta
 S = V\delta s$) is obtained in the form
\begin{equation}
 \delta s = - \frac{H_{co}^2}{4\pi T_{c0}}\left\{ \varphi^2_{eq} +
 \frac{C}{2}\left[\left(1+
\mu\varphi^2_{eq}\right)\mbox{ln}\left(1+\mu\varphi^2_{eq}\right) -
\mu\varphi^2_{eq}\mbox{ln}\left(\mu\varphi^2_{eq}\right)\right]\right\}\:.
\end{equation}

For the Al film of thickness $L_0 = 0.1\mu$m, the $C-$term in the curly
brackets of Eq.~(8) is of order $10^{-3}\varphi^2_{eq}$ and can be
neglected. Taking the value of the order parameter jump $\varphi_{eq}
\sim 0.033$ from Fig.~2, we obtain
\begin{equation}
\delta s = - \frac{H_{co}^2}{4\pi T_{c0}} \varphi^2_{eq} \approx \; -
0.7 \;\frac{\mbox{\footnotesize erg}}{\mbox{\footnotesize K.cm}^3}\:.
\end{equation}

In order to show the significance of this result we compare it with the
specific heat capacity jump at the second order phase transition point,
$T_{c0}$, described by the free energy $f_0(\varphi)$:
\begin{equation}
\delta C(T_{c0}) = \frac{H^2_{c0}}{4\pi T_{c0}} \approx 660 \;
\frac{\mbox{\footnotesize erg}}{\mbox{\footnotesize K.cm}^3}\:
\end{equation}

The ratio $(\delta s/\delta C) \sim 0.001$ is $10^{3}$ times bigger
than the respective quantity for bulk Al~\cite{HLM:1974}. Comparing
with the  results~\cite{TET:2002} for bulk Al one can see that the
metastability domain $(T^{\ast} - T_{eq})$ for  the film under
consideration is also 10$^3$ times larger than for bulk Al. The
calculated value ($\varphi_{eq} \approx 0.0032$) of the order parameter
jump in bulk Al is about 0.1 of that shown in Fig.~2. While the
overheated and stable superconducting states in quasi-2D Al films
appear only for $t_0 < 0$, i.e. below $T_{co}$, in 3D Al these states
occur also slightly above $T_{c0}$.

The present results are consistent with those obtained by Landau
expansion~\cite{FSU:2001} of the free energy~(6). However, there is  a
difference in the numerical values of some parameters considered in the
present investigation and those in Ref.~\cite{FSU:2001}. Obviously, the
Landau expansion has some limitations when  applied to free energies of
the type~(6).

{\bf 4. Conclusion}

The results show that Al films of thickness below $0.1\mu$m can be used
for an experimental test of the HLM effect. Calculations for  thinner
Al films or for W, where the GL parameter is 10 times smaller, will
give much stronger effect and, hence, better opportunity for an
experimental test. The thickness of the film may be lowered up to the
nm scale, below which the superconductivity is destroyed.

{\bf Acknowledgements.} Useful discussions with Profs. P. Fulde, R.
Klem, K. Maki, A. C. Mota, and Yu. N. Ovchinnikov, are gratefully
acknowledged. One of us (D.I.U.) thanks the hospitality of MPI-PKS
(Dresden).

\end{document}